\documentclass[twocolumn,aps,superscriptaddress,nofootinbib,floatfix]{revtex4}

\usepackage{epsfig,bm,feynmf}

\usepackage{graphics}

\usepackage[normalem]{ulem}  % \sout{old text} for strikeout
\usepackage[dvips]{color} % For blue in-text comments and additions

\renewcommand\sout{\bgroup \color{red} \ULdepth=-.5ex \ULset}
\begin{document}

\title{Particle production from gluon-nucleon interactions in relativistic heavy ion collisions}
\author{ Yong-Ping Fu,$^{1}$ Fei-Jie Huang,$^{2}$ and Qi-Hui Chen$^{3}$
\\ $^{1}$Department of Physics, West Yunnan University, Lincang 677000, China
\\ $^{2}$ Department of Physics, Kunming University, Kunming 650214, China
\\ $^{3}$ School of Physical Science and Technology, Southwest Jiaotong University, Chengdu 610031, China}
\date{\today}

\begin{abstract}
We propose a particle production mechanism analogous to the particle photoproduction processes, arising from the gluon-nucleon interactions in relativistic heavy ion collisions. The comparison is made on the effect of the gluon-nucleon interactions on the photon production in Au+Au collisions at $\sqrt{s_{NN}}=$200 GeV and Pb+Pb collisions at $\sqrt{s_{NN}}=$2.76 TeV. The numerical results indicate that as the collision energy increases, the contribution of gluon-nucleon interactions becomes more prominent.

\end{abstract}

%\pacs{xxx}

\maketitle

\section{Introduction}

Relativistic heavy ion collisions represent a prominent research focus in high energy physics. The high energy heavy ion collision experiments provide insights into various aspects of particle interactions, including the perturbative Quantum Chromodynamics (pQCD) that governs nucleon-nucleon deep inelastic scattering and their underlying parton structures. Moreover, the relativistic heavy ion collisions provide a unique opportunity to investigate the properties of hot and dense matter composed of quarks and gluons, known as quark-gluon plasma (QGP).

In relativistic heavy ion collisions, the jets produced from the parton collisions lose energy due to gluon radiation in the hot medium, this phenomenon is called jet quenching \cite{WHS-JQ1,WHS-JQ2,WHS-JQ3}. Jets in a hot medium can radiate gluons similarly to the bremsstrahlung of electrons. In the color deconfined QGP, jet quarks with color charge radiate soft gluons due to multiple scattering with the hot particles \cite{BDMPS1,BDMPS2,GW1,GW2}. The suppression of hadron spectra from relativistic heavy ion collisions observed by detectors at Relativistic Heavy Ion Collider (RHIC) and Large Hadron Collider (LHC) can provide evidences for the existence of this effect \cite{JQ-ex1,JQ-ex2,JQ-ex3}. Without the influence of the thermal medium, partons within nucleons can also radiate gluons through hard scattering \cite{HS-g1,HS-g2,HS-g3,HS-g4}. However, the gluon radiation discussed above occurs in the final states of particle scattering. Due to color confinement, it is not possible for initial state partons within nucleons to directly radiate gluons in hadronic collisions.

\begin{figure}
\includegraphics[width=0.5\linewidth]{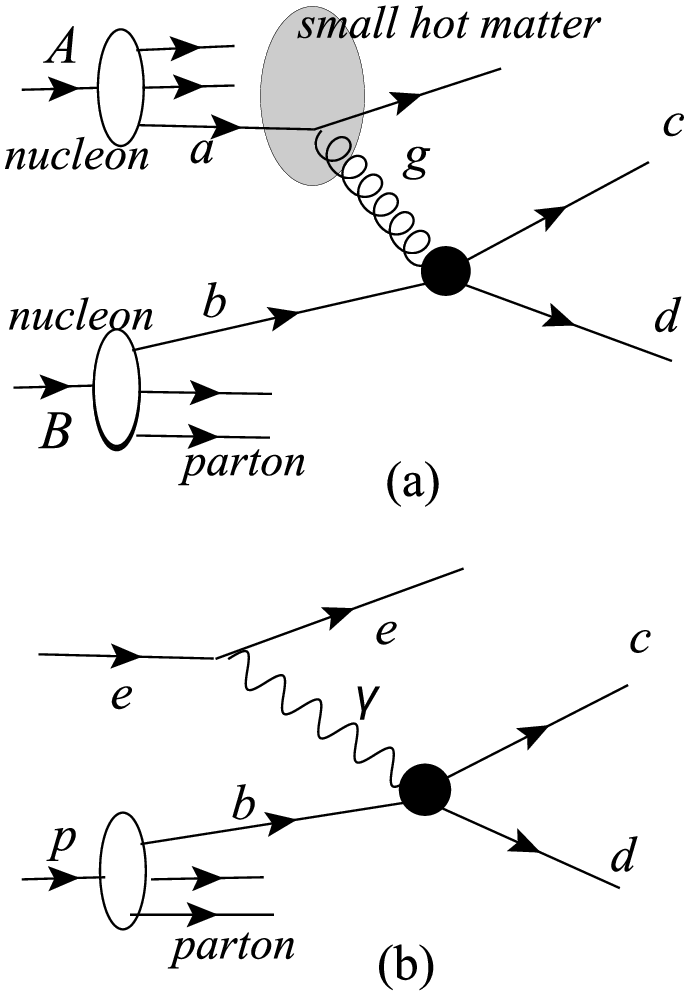}
\caption{\label{fig1} (a) Illustration of gluon-nucleon interactions, the gluon radiates from the initial state parton $a$ due to scattering with the small hot medium. (b) The $e$-$p$ deep inelastic scattering induced by the photon-nucleon interactions.}
\end{figure}

The collective behavior of hadronic particles has been observed in high multiplicity $pp$, $pA$ and $dA$ collision experiments at RHIC and LHC \cite{smallQGP01,smallQGP02,smallQGP03,smallQGP04,smallQGP1,smallQGP2,smallQGP3,smallQGP4,smallQGP5}. These experiments indicate that in the early stages of relativistic heavy ion collisions, some nucleon collisions lead to localized thermalization, resulting in the formation of small hot systems. In this paper, we consider the situation in which some nucleons of heavy ions have already collided and thermalized in the early stage of the relativistic heavy ion collisions. The remaining nucleons (or partons within the nucleons) in the heavy ion, which have not yet collided, will collide and interact with these small hot systems. If the mean free path of the incident parton is large compared to the Debye screened length, an initial state quark $a$ of the nucleon $A$ collides into the small hot medium, and scatters with a thermal parton, the initial state quark $a$ will radiate a high energy gluon. The radiated gluon will then interact with an parton $b$ of another incident nucleon $B$, the process [Fig.\ref{fig1}(a)] is similar to the particle photoproduction processes in electron-proton deep inelastic scattering at HERA \cite{HERA} [Fig.\ref{fig1}(b)].

Through the above analysis, we can observe that under the condition of color deconfinement, the high energy gluons radiated by initial state quarks are entirely capable of undergoing hard scattering with other cold partons. This potential gluon-nucleon interaction in the early stages of relativistic heavy ion collisions will become a new candidate mechanism for the particle production.

%\begin{figure*}
%\includegraphics[scale=0.4]{Graph01}
%\end{figure*}

The paper is organized as follows: In Sec. II, we discuss the scattering amplitude for the quark scattering that leads to the gluon radiation. The gluon spectrum inside an initial state quark is calculated. In Sec. III, we derive the differential cross section for particles production resulting from the gluon-nucleon interactions in relativistic heavy ion collisions. In Sec. IV, we employ photon probes to analyze the contribution of the gluon-nucleon interactions at RHIC and LHC energies. Finally, a summary is provided in Sec. V.

\section{Gluon radiation of the initial state quark}

\begin{figure}
\includegraphics[width=0.8\linewidth]{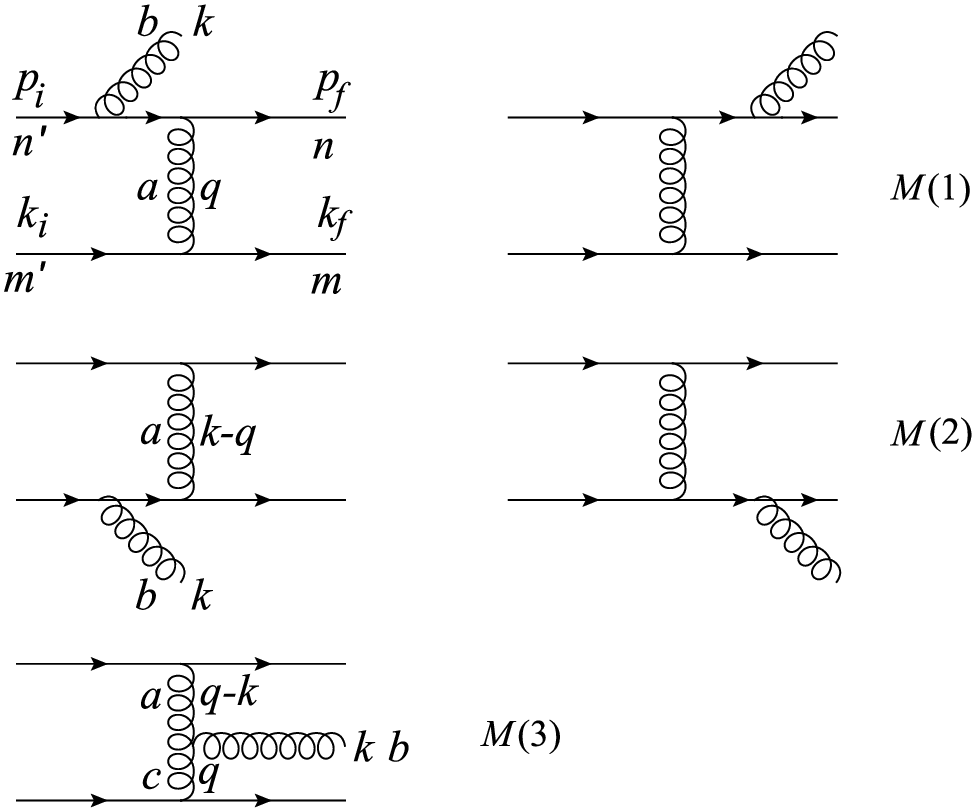}
\caption{\label{fig2} Feynman diagrams for gluon radiation in leading order (LO) quark scattering.}
\end{figure}

We first review the amplitude of quark scattering that leads to gluon radiation \cite{GW2,HS-g1}, the Feynman diagrams satisfying gauge invariance are shown in Fig.\ref{fig2}. The complete LO amplitude of gluon radiation induced by the quark scattering is given by
\begin{eqnarray}\label{amplitude1}
M\!\!&=&\!\!\frac{ig^{3}}{q^{2}}\!\!  \bar{u}(p_{f})\gamma_{\mu}u(p_{i})\bar{u}(k_{f})\gamma^{\mu}u(k_{i})  \nonumber\\[1mm]
&& \left\{ \left[\!\frac{\varepsilon\!\cdot \!p_{f}}{k\!\cdot \! p_{f}}\!\!\left(\!T^{b}T^{a}\!\right)_{nn'} \!-\! \frac{\varepsilon\!\cdot\! p_{i}}{k\!\cdot \! p_{i}}\!\!\left(\!T^{a}T^{b}\!\right)_{nn'}\!\right]\!\!T^{a}_{mm'} \right.  \nonumber\\[1mm]
&& +        \frac{q^{2}}{(k-q)^{2}}\!\!\left[\!\frac{\varepsilon\!\cdot \!k_{f}}{k\!\cdot \! k_{f}}\!\!\left(\!T^{b}T^{a}\!\right)_{mm'} \!-\! \frac{\varepsilon\!\cdot\! k_{i}}{k\!\cdot \! k_{i}}\!\!\left(\!T^{a}T^{b}\!\right)_{mm'}\!\right]\!\!T^{a}_{nn'}  \nonumber\\[1mm]
&& +  \left. \frac{2\varepsilon\cdot q}{(k-q)^{2}}[T^{b},T^{a}]_{nn'}T^{a}_{mm'}  \right\},
\end{eqnarray}
where $T^{a}$ is the SU(3) generator with the color indices $n$, $n'$, $m$, and $m'$ of quarks. Here $a$ and $b$ are the non-Abelian gauge field indices. The three terms of the above equation represent the scattering amplitudes $M(1)$, $M(2)$ and $M(3)$ in the Feynman diagrams as shown in Fig.\ref{fig2}, respectively. The radiated gluon momentum $k$ satisfies the on-shell condition $k^{2}$=0 , and the polarization $\varepsilon$ satisfies $\varepsilon\cdot k$=0.

\begin{figure}
\includegraphics[width=0.5\linewidth]{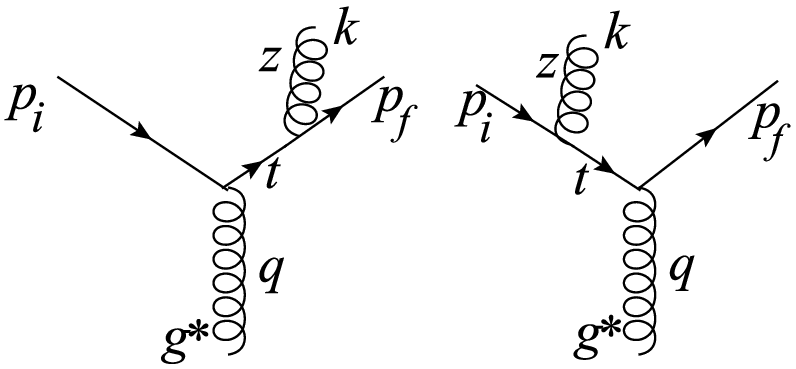}
\caption{\label{fig3} The amplitudes for the LO gluon radiation.}
\end{figure}

The light-cone initial momenta $p_{i}$ of the beam parton in the nucleon and $k_{i}$ of the target parton in the small hot medium are defined as \cite{GW2}
\begin{eqnarray}\label{pi}
p_{i}=(p^{+},0,\mathbf{0}_{\bot}),
\end{eqnarray}
\begin{eqnarray}\label{ki}
k_{i}=(k_{i}^{+},k_{i}^{-},\mathbf{0}_{\bot}),
\end{eqnarray}
where the thermal momentum $k_{i}^{+}=k_{i}^{-}\sim T$ in the QGP rest frame. The final momenta of the beam and target partons are
\begin{eqnarray}\label{pf}
p_{f}=p_{i}+q-k,
\end{eqnarray}
\begin{eqnarray}\label{kf}
k_{f}=k_{i}-q.
\end{eqnarray}
The GW and BDMPS models have considered the effect of soft gluon radiation, and found that the amplitude $M(2)$ in Fig.\ref{fig2} is negligible compared to $M(1)$ and $M(3)$ in the light-cone gauge \cite{GW1,GW2,BDMPS1,BDMPS2}. Here we show that in the high energy gluon radiation limit, $k \gg q$, the amplitude $M(2)$ and $M(3)$ are depressed by the terms of $q^{2}/(k-q)^{2}$ and $\varepsilon\cdot q/(k-q)^{2}$, respectively. Only the projectile diagrams $M(1)$ contribute significantly to the radiation of high energy gluon. In the leading pole approximation the differential cross section of the gluon radiation (Fig.\ref{fig3}) can be expressed as \cite{Field}

\begin{eqnarray}\label{crosssection}
\frac{d\sigma}{\sigma_{0}dzdt}=\frac{\alpha_{s}}{2\pi t}P_{gq}(z),
\end{eqnarray}
where the Mandelstam variable $t=(p_{i}+q)^{2}=(k+p_{f})^{2}$, the splitting function $P_{gq}=C_{F}[1+(1-z)^{2}]/z$ with the Casimir $C_{F}=4/3$. $z$ is the momentum fraction of the radiated gluon. By integrating over the Lorentz invariant $t$, we obtain the gluon radiation spectrum of an initial state quark
\begin{eqnarray}\label{fg}
f_{gq}=\int dt\frac{d\sigma}{\sigma_{0}dzdt}=\frac{\alpha_{s}}{2\pi }P_{gq}(z)\ln\frac{t_{max}}{t_{min}},
\end{eqnarray}
the values of $t_{max}$ and $t_{min}$ will discuss in Section III.

\section{Particles production induced by gluon-nucleon interactions}

The invariant differential cross section for particle production from the gluon-nucleon interactions in Fig.\ref{fig1}(a) can be derived as following
\begin{eqnarray}\label{cross1}
\frac{d\sigma_{g\!-\!n}}{d^{2}p_{T}dy}&\!\!\!\!=\!\!\!&\sum_{a,b}\!\frac{1}{\pi}\!
\!\!\int_{x_{a}^{min}}^{1}\!\!dx_{a}\!\!\int_{x_{b}^{min}}^{1}\!\!dx_{b}G_{a}^{A}(x_{a},Q^{2})f_{ga}(z_{a},T) \nonumber\\[1mm]
&& \times G_{b}^{B}(x_{b},\!Q^{2})\frac{x_{a}x_{b}z_{a}}{x_{a}x_{b}-x_{a}x_{2}}\frac{d\hat{\sigma}_{gb\rightarrow c d}}{d\hat{t}}(\hat{s},\!\hat{u},\!\hat{t}),
\end{eqnarray}
where $x_{a,b}$ and $z_{a}$ are the momentum fractions of the initial state parton and radiated gluon, respectively. Here the subscripts $a$, $b$, and $c$ represent the partons. $x_{2}=p_{T}e^{-y}/\sqrt{s_{NN}}$, $\sqrt{s_{NN}}$ is the center-of-mass
energy of the colliding nucleons, $Q$ is the momentum scale. The momentum fractions with the transverse momentum $p_{T}$ and rapidity $y$ are given by
\begin{eqnarray}\label{xamin}
x_{a}^{min}=\frac{p_{T}e^{y}}{\sqrt{s_{NN}}-p_{T}e^{-y}},
\end{eqnarray}
\begin{eqnarray}\label{xbmin}
x_{b}^{min}=\frac{x_{a}p_{T}e^{-y}}{x_{a}\sqrt{s_{NN}}-p_{T}e^{y}},
\end{eqnarray}
\begin{eqnarray}\label{z}
z_{a}=\frac{x_{b}p_{T}e^{y}}{x_{a}x_{b}\sqrt{s_{NN}}-x_{a}p_{T}e^{-y}}.
\end{eqnarray}
The parton distribution for the
nucleus is given by
\begin{eqnarray}\label{PDF}
G_{a}^{A}(x_{i},Q^{2})\!\!\!&=&\!\!\!R^{A}_{a}(x_{i},Q^{2})[Z F_{a}^{p}(x_{i},Q^{2})  \nonumber\\[1mm]
&&  +(A-Z)F_{a}^{n}(x_{i},Q^{2})]/A,
\end{eqnarray}
where $R^{A}_{a}(x_{i},Q^{2})$ is the nuclear modification factor, $Z$ is the proton number, $A$ is the nucleon number. The functions $F_{a}^{p}(x_{i},Q^{2})$ and
$F_{a}^{n}(x_{i},Q^{2})$ are the parton distribution function (PDF) of the proton
and neutron, respectively.

The Mandelstam variables $\hat{s}$, $\hat{u}$ and $\hat{t}$ of the differential cross sections $d\hat{\sigma}/d\hat{t}(gb\rightarrow cd)$ are
\begin{eqnarray}\label{s}
\hat{s}=x_{a}x_{b}z_{a} s_{NN},
\end{eqnarray}
\begin{eqnarray}\label{u}
\hat{u}=-x_{b}p_{T}e^{y}\sqrt{s_{NN}},
\end{eqnarray}
\begin{eqnarray}\label{t}
\hat{t}=-x_{a}z_{a} p_{T}e^{-y}\sqrt{s_{NN}}.
\end{eqnarray}

In the high energy gluon radiation limit, $q\ll k$, the logarithmic scale $t_{max}$ in Eq.(\ref{fg}) is $t_{max}=(p_{imax}+q_{max})^{2}$, here $p_{imax}\approx (\sqrt{\hat{s}},0,\mathbf{0}_{\perp})$, $q_{max}\approx (T,T,\mathbf{0}_{\perp})$. In this case, we obtain the results
\begin{eqnarray}\label{tmax}
t_{max}=\left(\sqrt{\hat{s}}+T\right)T.
\end{eqnarray}
Taking into account the scale parameter of pQCD, we choose $t_{min}$=$\Lambda^{2}$ \cite{th r 3}. The parameter $t_{max}$ indicates that the gluon radiation spectrum depends on the temperature of the QGP droplets formed in the early stage of relativistic heavy ion collisions.

\section{Application in Photon production}

As an application of gluon-nucleon scattering, we will now discuss particles that can be detected by the detectors, such as photons. The electromagnetic radiation produced from relativistic heavy ion collisions is a useful probe for investigating pQCD and the QGP. Photons do not directly participate in strong interactions, and the mean free path of photons is larger than that of the collision system. Consequently, photons can escape to the detector almost undistorted through the strongly interacting system.

In the discussion of photon production, the leading-order (LO) subprocess of gluon-nucleon collision is the Compton scattering $gq\rightarrow \gamma q$ in Eq.(\ref{cross1}). In relativistic heavy ion collisions, the conventional sources of hard photon production include direct photons and fragmentation photons. The direct photons are primarily produced by the quark-antiquark annihilation ($q\bar{q}\rightarrow g\gamma$) and the Compton scattering ($qg\rightarrow q
\gamma$) \cite{th r 3}. The invariant cross section from the two processes is expressed as
%\begin{widetext}
\begin{eqnarray}\label{dir}
\frac{d\sigma_{dir\gamma}}{d^{2}p_{T}dy}&=&\sum_{a,b}\frac{1}{\pi}\int^{1}_{x^{min}_{a}}
dx_{a}G_{a}^{A}(x_{a},Q^{2})G_{b}^{B}(x_{b},Q^{2})
\nonumber\\[1mm]
&&\times
\frac{x_{a}x_{b}}{x_{a}-x_{1}}\frac{d\hat{\sigma}_{ab\rightarrow \gamma d}}{d\hat{t}}(\hat{s},\hat{u},\hat{t}),
\end{eqnarray}
%\end{widetext}
here $x_{1}=p_{T}e^{y}/\sqrt{s_{NN}}$, the momentum fraction $x_{b}$ is
\begin{eqnarray}
x_{b}=\frac{x_{a}p_{T}e^{-y}}{x_{a}\sqrt{s_{NN}}-p_{T}e^{y}},
\end{eqnarray}
and the value of $x_{a}^{min}$ is consistent with Eq.(\ref{xamin}). The Mandelstam variables of the the subprocesses $d\hat{\sigma}/d\hat{t}(ab\rightarrow \gamma d)$ are
\begin{eqnarray}
\hat{s}=x_{a}x_{b}s_{NN},
\end{eqnarray}
\begin{eqnarray}
\hat{u}=-x_{b}p_{T}e^{y}\sqrt{s_{NN}},
\end{eqnarray}
\begin{eqnarray}
\hat{t}=-x_{a} p_{T}e^{-y}\sqrt{s_{NN}}.
\end{eqnarray}

In addition to direct photon production, fragmentation photons are also the primary source of large transverse-momentum photons in initial parton collisions \cite{th r 3}. The invariant cross section for the $AB\rightarrow (c\rightarrow\gamma)X$ interaction can be written as
\begin{eqnarray}\label{frag}
\frac{d\sigma_{fra\gamma}}{d^{2}p_{T}dy}&\!\!=&\!\!\sum_{a,b}\frac{1}{\pi}\int^{1}_{x_{a}^{min}}\!\!
\!\!dx_{a}\int^{1}_{x_{b}^{min}}\!\! \!\!dx_{b} G_{a}^{A}(x_{a},Q^{2}) G_{b}^{B}(x_{b},Q^{2})
\nonumber\\[1mm]
&&\times
D_{\gamma c}(z_{c},Q^{2})\frac{1}{ z_{c}}
\frac{d\hat{\sigma}_{ab\rightarrow c d}}{d\hat{t}}(\hat{s},\hat{u},\hat{t}),
\end{eqnarray} %\end{widetext}
where the momentum fraction of the final state parton $c$ is
\begin{eqnarray}
z_{c}=\frac{p_{T}}{\sqrt{s_{NN}}}\left(\frac{e^{y}}{x_{a}}+\frac{e^{-y}}{x_{b}}\right).
\end{eqnarray}
The values of $x_{a}^{min}$ and $x_{b}^{min}$ are consistent with Eq.(\ref{xamin}) and Eq.(\ref{xbmin}), respectively. The Mandelstam variables of the differential cross sections $d\hat{\sigma}/d\hat{t}(ab\rightarrow cd)$ are
\begin{eqnarray}
\hat{s}=x_{a}x_{b}s_{NN},
\end{eqnarray}
\begin{eqnarray}
\hat{u}=-\frac{x_{b}}{z_{c}}p_{T}e^{y}\sqrt{s_{NN}},
\end{eqnarray}
\begin{eqnarray}
\hat{t}=-\frac{x_{a}}{z_{c}} p_{T}e^{-y}\sqrt{s_{NN}}.
\end{eqnarray}

\begin{figure}[t]
\includegraphics[width=1.0\linewidth]{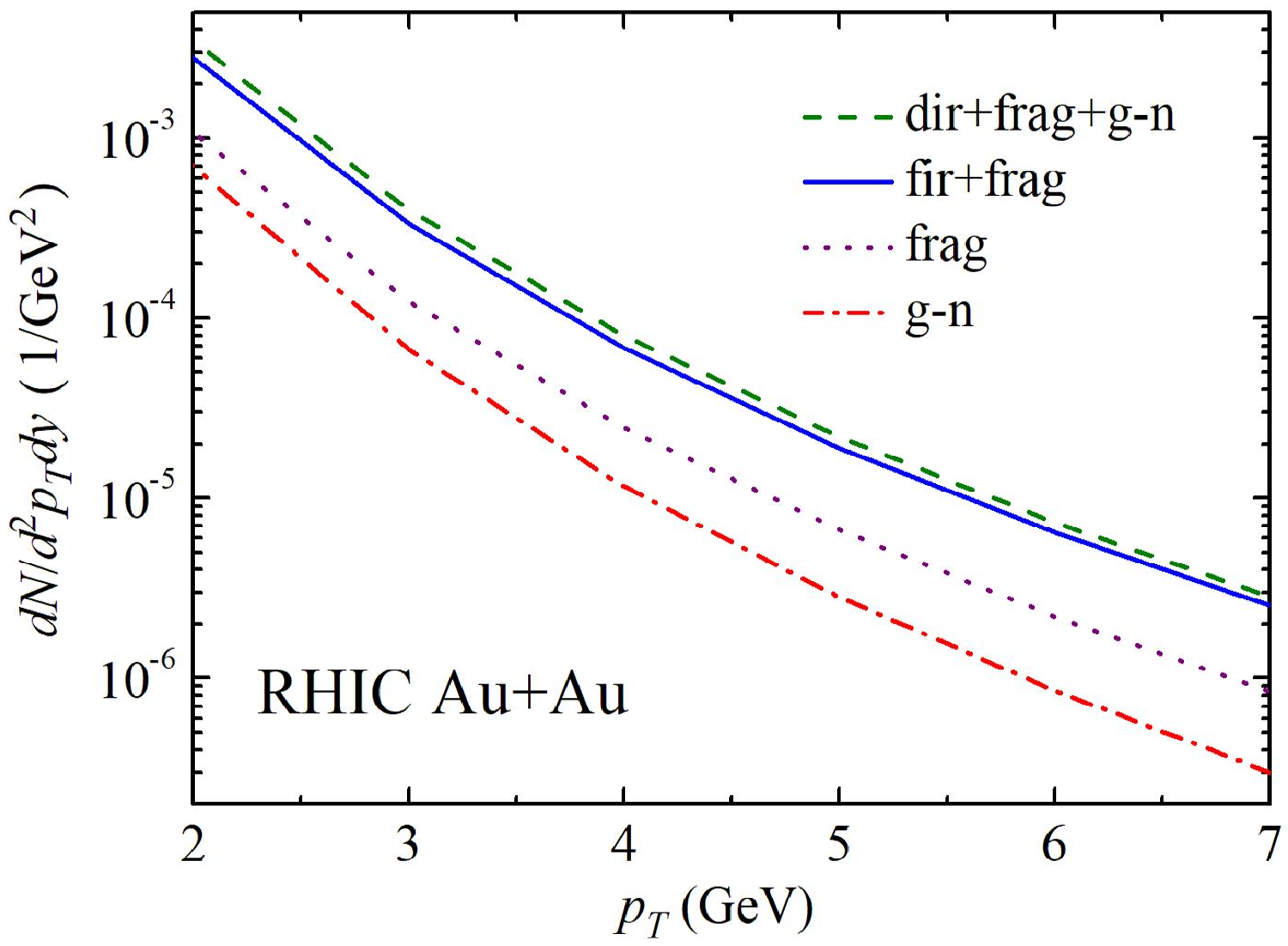}
\caption{\label{fig4} The large transverse momentum photons produced from the Au+Au collisions at $\sqrt{s_{NN}}$=200 GeV. The solid line is the photon yield from the direct photons and fragmentation photons. The dashed line includes the direct photons, fragmentation photons and photons produced from the gluon-nucleon interactions. The dash dot line is the photon yield from the gluon-nucleon interactions.}
\end{figure}

The effect of thermal medium induced jet energy loss on photon fragmentation $D^{0}_{\gamma c}$ is presented by the WHS phenomenological model \cite{WHS-JQ1,WHS-JQ2,WHS-JQ3}, this approach is useful for studies of the parton energy loss of fragmentation function and multiple final state scattering. The probability for a jet to scatter $n$ times within a distance $L(=n\lambda_{q})$ in the thermal medium is provided as $P_{n}=(L/\lambda_{q})^{n}e^{- L/\lambda_{q}}/n!$,
where $\lambda_{q}$ is the mean free path of the quark. The fragmentation function can be modified in the following form
\begin{eqnarray}
D_{\gamma c}(z_{c},Q^{2})=C_{n}\sum^{N}_{n=0}P_{n}\frac{z_{c}^{n}}{z_{c}}D_{\gamma c}^{0}(z^{n}_{c},Q^{2})
\end{eqnarray}
where $C_{n}=1/\left(\sum^{N}_{n=0}P_{n}\right)$, $N=E^{\mathrm{jet}}_{T}/\varepsilon$ is the scattering number, $E^{\mathrm{jet}}_{T}=p_{T}/z_{c}$ is the transverse energy of the jet, and $z^{n}_{c}=z_{c}/\left(1-\Delta E/E^{\mathrm{jet}}_{T} \right)$. The total energy loss $\triangle E$ of the quark and average energy loss per scattering $\varepsilon$ are defined as follows:
\begin{eqnarray}
\triangle E=\int_{0}^{L}\frac{dE}{dx} dx,
\end{eqnarray}
\begin{eqnarray}
\varepsilon=\lambda_{q}\frac{dE}{dx}.
\end{eqnarray}
The energy loss $dE/dx$ of jets crossing the hot and dense plasma is determined by the BDMPS model \cite{BDMPS1}
\begin{eqnarray}
\frac{dE}{dx}=\frac{\alpha_{s}c_{a}\mu^{2}}{8\lambda_{g}}L\ln\frac{L}{\lambda_{g}},
\end{eqnarray}
where $c_{a}$=4/3 for quarks and 3 for gluon, the square of the Debye mass is $\mu^{2}=4\pi\alpha_{s}T^{2}$, $\lambda_{g}=\pi\mu^{2}/\left[126\alpha_{s}^{2}\zeta(3)T^{3}\right]$ and $\lambda_{q}=9\lambda_{g}/4$ are the gluon and quark mean free path, respectively \cite{GW2}.

\begin{figure}[t]
\includegraphics[width=1.0\linewidth]{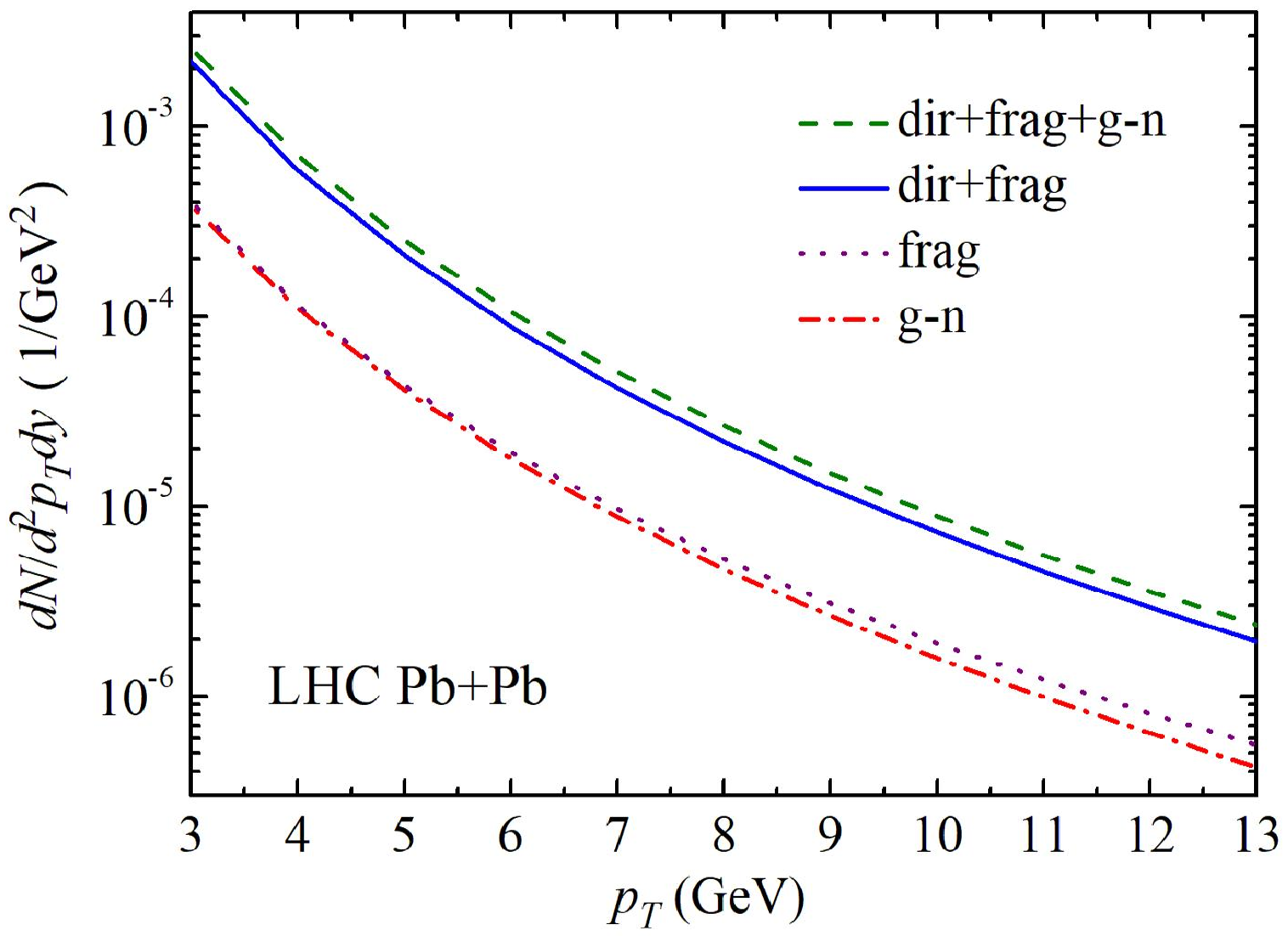}
\caption{\label{fig5} Same as Fig.4 but for the Pb+Pb collisions at $\sqrt{s_{NN}}$=2.76 TeV. }
\end{figure}

In Figs.\ref{fig4} and \ref{fig5}, we present the yields of direct photons, fragmentation photons, and photons originating from gluon-nucleon interactions in Au+Au collisions at $\sqrt{s_{NN}}$=200 GeV and Pb+Pb collisions at $\sqrt{s_{NN}}$=2.76 TeV. The contribution of gluon-nucleon interactions in Au+Au collisions at RHIC is not as pronounced as in Pb+Pb collisions at the LHC. Through the analysis of photon production processes, we observed that, in comparison to the invariant differential cross section of direct photon production, the contribution of the gluon-nucleon scattering processes is suppressed by the QCD running coupling constant $\alpha_{s}$ in the gluon radiation spectral function $f_{gq}$. However, as the collision energy increases, we find that the correction contribution of gluon-nucleon interactions to the production of hard photons also strengthens. Through numerical calculations, we find that at the RHIC energy, the contribution of gluon-nucleon (g-n) interactions to the yield of hard photons (dir+frag) shows a decreasing trend, dropping from 25\% at $p_{T}$=2 GeV to 12\% at $p_{T}$=7 GeV. At the LHC energy, however, the contribution rate of gluon-nucleon interactions shows an increasing trend, rising gradually from 17\% at $p_{T}$=3 GeV to 22\% at $p_{T}$=13 GeV.

Although the invariant cross section magnitude for gluon-nucleon collisions [$O(\alpha_{s}^{2}\alpha)$] is smaller than the magnitude for direct photon production [$O(\alpha_{s}\alpha)$], we note that the invariant cross section for fragmentation photons [Eq.(\ref{frag})] is of the same order as gluon-nucleon collisions [Eq.(\ref{cross1})]. There is only one LO subprocess for gluon-nucleon interactions, while there are at least six LO subprocesses for fragmentation photon production. However, we still observe that in Pb+Pb collisions at 2.76 TeV, the photon yield from gluon-nucleon interactions is essentially comparable to the yield of fragmentation photons. (Fig.\ref{fig5}).

The photon yield is obtained by the following \cite{jet-r 1}
\begin{eqnarray}
\frac{dN}{d^{2}p_{T}dy}=\frac{\langle N_{coll}\rangle}{\sigma_{inel}^{NN}}\frac{d\sigma}{d^{2}p_{T}dy},
\end{eqnarray}
where $\langle N_{coll}\rangle$ is the average number of binary nucleon-nucleon collisions, $\sigma_{inel}^{NN}$ is the inelastic nucleon-nucleon cross section. We use $\langle N_{coll}\rangle$=770.6 and 1210.8, $\sigma_{inel}^{NN}$=40 and 64 mb, corresponding to the nucleon-nucleon collisions at $\sqrt{s_{NN}}$=200 GeV and 2.76 TeV, respectively \cite{r ex1,r ex2,jet-r 1,r ex3,Ncoll1,Ncoll2}. In the numerical calculations, we use the CTEQ6L1 PDF \cite{CTEQ6M 1} and EPS09 nuclear modifications \cite{EPS09 1}. The photon productions are calculated in the midrapidity region. The photon fragmentation function is used by Owens' parameterization results \cite{th r 3}. The correction factors $K_{dir}\sim$1.5 for RHIC and LHC, $K_{frag}\sim$1.8 at RHIC and 1.4 at LHC are used to account for the next-to-leading order (NLO) corrections \cite{jet-r 1}.

\begin{figure}[t]
\includegraphics[width=1.0\linewidth]{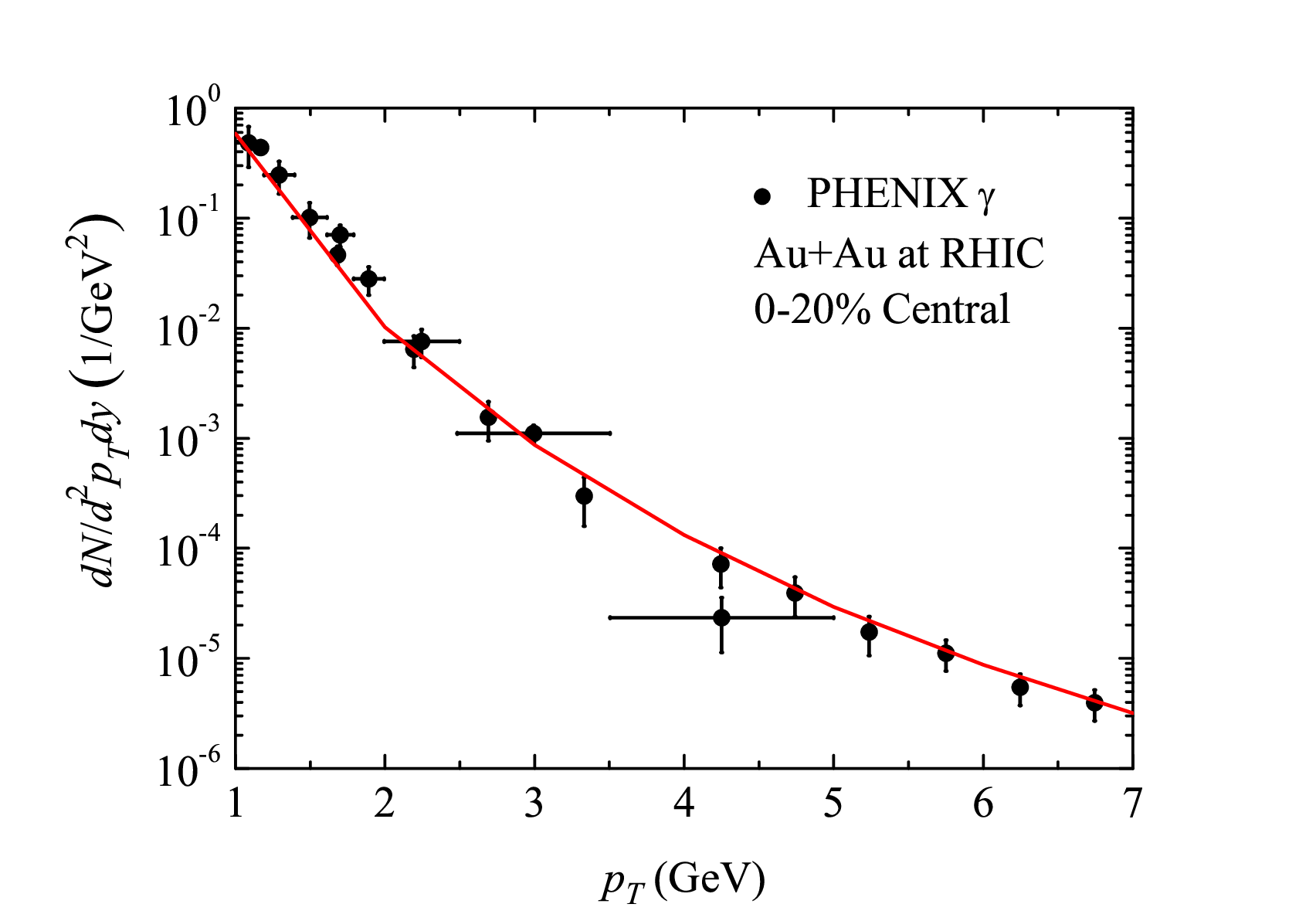}
\caption{\label{fig6}  Photon yields from the Au+Au collisions at $\sqrt{s_{NN}}$=200 GeV for 0-20$\%$ centrality class. The solid line includes the sum of direct photons, fragmentation photons, jet-photon conversion, thermal photons from the QGP and HG phases, and photons produced from the gluon-nucleon interactions. The data on photons are from the PHENIX experiments \cite{r ex1,r ex2}.}
\end{figure}

\begin{figure}[t]
\includegraphics[width=1.0\linewidth]{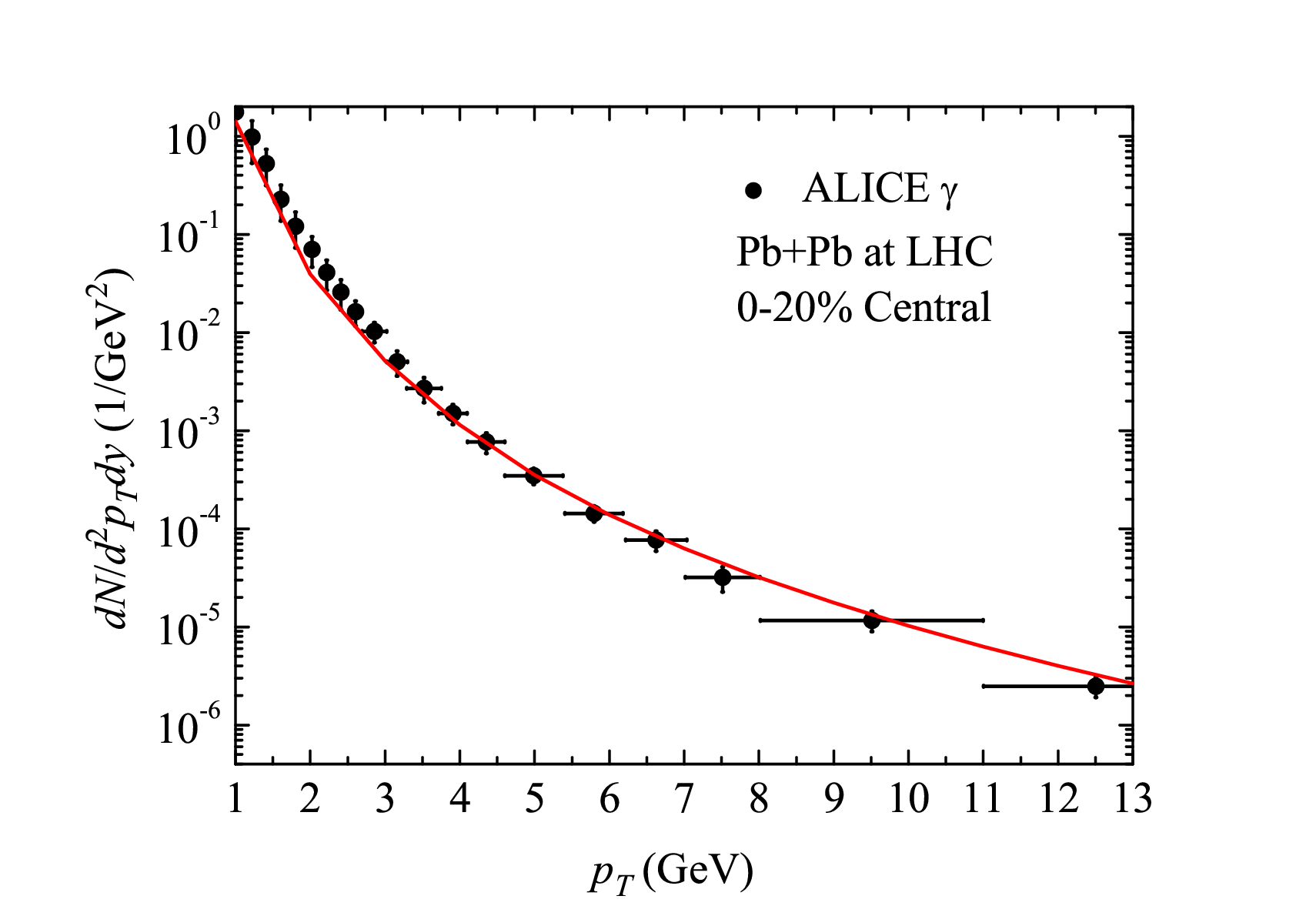}
\caption{\label{fig7} Same as Fig.6 but for the Pb+Pb collisions at $\sqrt{s_{NN}}$=2.76 TeV for 0-20$\%$ centrality class. The data on photons are from the ALICE experiments \cite{r ex3}.}
\end{figure}

Hard photons are mainly produced by the interactions of the initial partons discussed above. We mainly focus on the production of large transverse momentum photons. Since thermal photons are primarily concentrated in the low transverse momentum region, the photons produced from interactions of the cold components are not prominent in this region \cite{th r 11,1+1D 2}. Thermal photons emitted from the QGP \cite{th r 1,th r 2,th r 4,th r 5,th r 6,th r 7,th r 8,th r 9,th r 10,th r 11} and Hadronic Gas (HG) \cite{th r 9,HG 2,HG 3,HG 4} phases are mainly concentrated in the low transverse momentum region. Jets from cold component interactions colliding with the hot medium can also produce large transverse momentum photons \cite{jet-r 1,jet-r 2,jet-r 3,jet-r 4,jet-r 5}. In Figs.\ref{fig6} and \ref{fig7}, we present the total photon yield from relativistic heavy ion collisions at RHIC and LHC energies. Numerical calculations indicate that the photon contribution from gluon-nucleon interactions does not exceed the experimental data error range. Due to the limitations of electromagnetic signal detectors at RHIC and LHC, isolating photons from gluon-nucleon interactions in relativistic heavy ion collision experiments is not feasible. However, by using the photon probe, we discussed the possibility of high energy gluon radiation by initial partons in relativistic heavy ion collisions.

\section{Summary}
In this work we present calculations of the temperature dependent spectrum of high energy gluon radiation and the invariant differential cross section for particle production from the gluon-nucleon interactions in relativistic heavy ion collisions. Through calculations of photon production at RHIC and LHC energies, we demonstrate that in the large transverse momentum region, gluon-nucleon interactions make a noticeable contribution to particle production. This contribution of gluon-nucleon interactions is directly related to the nucleon-nucleon collision energy, and it becomes more pronounced with increasing collision energy, particularly in the LHC energies. In relativistic heavy ion collisions, the gluon-nucleon interactions can be one of the candidate mechanisms for particle production.

\section*{Acknowledgements}
This work was supported by the Program for Innovative Research Team at Kunming University, the Program for Frontier Research Team at Kunming University 2023. Y. P. Fu acknowledges the support from the National Natural Science Foundation of China (Grant No.11805029).



%% References


\bibliographystyle{model1-num-names}
\bibliography{<your-bib-database>}

%\end{narrowtext}

\end{document}